# Investigating Student Difficulties with Time dependence of Expectation Values in Quantum Mechanics

Emily Marshman and Chandralekha Singh

*Department of Physics and Astronomy, University of Pittsburgh, Pittsburgh, PA, 15260, USA*

**Abstract:** Quantum mechanics is challenging even for advanced undergraduate and graduate students. In the Schrödinger representation, the wave function evolves in time according to the time dependent Schrödinger equation. The time dependence of the wave function gives rise to time dependence of the expectation value of observables. We have been exploring the difficulties that advanced undergraduate and graduate students have with time dependence of expectation values in quantum mechanics. We have developed and administered conceptual free response and multiple-choice questions to students to investigate these difficulties. We also interviewed 23 students individually using a think-aloud protocol to obtain a better understanding of the rationale behind students' written responses. We find that many students struggle with time dependence of expectation values of observables. We discuss some findings.

## INTRODUCTION

Quantum mechanics concepts can be difficult to grasp, especially because many concepts are counter-intuitive [1-6]. We have been examining student difficulties in learning quantum mechanics. Here, we focus on an investigation to identify students' difficulties with concepts related to the time dependence of the expectation value of observables. This investigation was conducted with students at the University of Pittsburgh (Pitt) by administering written conceptual free response and multiple-choice questions as a part of upper-level undergraduate or graduate level courses and by conducting in-depth individual interviews with 23 students using a think-aloud protocol. In these semi-structured interviews, students were asked to talk aloud while they answered questions posed to them. The interviewer did not disturb students' thought processes while they answered the questions except encouraging them to keep talking if they became quiet. Later, the interviewer asked students for clarification of points they had not made clear earlier in order to understand their thought processes better. Some of these questions were planned out ahead of time while others were emergent queries based upon a particular student's responses during an interview.

Here, we present some results pertaining to the difficulties that first year physics graduate students have with the concept of time dependence of expectation value of operators corresponding to physical observables as manifested by their responses to conceptual multiple-choice questions administered in two different graduate classes. These questions were posed after eleven weeks of first semester graduate quantum mechanics for one set of graduate students (15 total number) and almost two full semesters of graduate quantum mechanics courses for the second set of graduate students (24 total number). We found that there is no significant difference between the scores of the two classes on the multiple choice questions. We also discuss the findings from think-aloud interviews conducted with individual graduate students to understand their difficulties in-depth immediately after they had successfully completed a sequence of two first year graduate quantum mechanics courses.

In all of the discussions below, we will assume that none of the operators corresponding to the physical observables in question (including the Hamiltonian of the system) have explicit time dependence. This simplifying constraint was explicitly mentioned in all of the questions students were asked. The general equation for the time dependence of the expectation value of an operator $\hat{Q}$ with no explicit time dependence in a state $|\psi\rangle$ is given by $\frac{d}{dt}\langle\hat{Q}\rangle = \frac{i}{\hbar}\langle[\hat{H},\hat{Q}]\rangle$, where $\langle[\hat{H},\hat{Q}]\rangle$ is the commutator of the Hamiltonian operator $\hat{H}$ and the operator $\hat{Q}$. Students must conceptually understand that there are two conditions that can be derived from this general equation for when the expectation value of an operator $\hat{Q}$ corresponding to an observable does not depend on time. In particular, the expectation value of an operator that commutes with the Hamiltonian is time independent regardless of the initial state. Moreover, if the system is initially in an eigenstate of the Hamiltonian or energy eigenstate, the expectation value of any operator (no explicit time dependence) will be time independent. In the course of these investigations, we find that many advanced students struggle with these issues related to the time dependence of expectation values.

## STUDENT DIFFICULTIES

One source of difficulty with concepts related to the time dependence of the expectation value of an observable was that students struggled with the time dependence of the wave function. Many students did not understand that the Hamiltonian of the system plays a crucial role in the time evolution of the state. The eigenstates of the Hamiltonian are special pertaining to time development issues for a quantum system. In fact, this central role of the Hamiltonian in the time development of the system is responsible for the energy eigenstates also being called the stationary states of the system. In a stationary state, the expectation value of any observable with no explicit time dependence will not depend on time.

Below, we discuss the performance of 39 students (combining the performances from two different years since the performances are similar) on four conceptual multiple choice questions given as part of conceptual surveys in their first year physics graduate courses and also discuss findings from individual interviews with a subset of the graduate students. Table 1 shows that approximately 50% or less of the graduate students answered these questions correctly.

**TABLE 1.** Percentages of physics graduate students who selected different answer choices on the four multiple choice questions about the time dependence of expectation values. (Correct answers are italicized.)

|  | A | B | C | D | E |
|---|---|---|---|---|---|
| Question 1 | 18 | 33 | 28 | 0 | *21* |
| Question 2 | *46* | 10 | 8 | 26 | 8 |
| Question 3 | 8 | 0 | 3 | *51* | 36 |
| Question 4 | *33* | 3 | 0 | 8 | 51 |

The following question probes students' understanding of the time dependence of the expectation value of an operator.

1. Choose all of the following statements that are correct about the time dependence of the expectation value of an observable Q in a state $|\psi\rangle$. Neither the Hamiltonian $\hat{H}$ nor the operator $\hat{Q}$ depends explicitly on time.

(1) $\quad \frac{d}{dt}\langle\hat{Q}\rangle = \frac{i}{\hbar}\langle[\hat{H},\hat{Q}]\rangle$.

(2) $\quad \frac{d}{dt}\langle\hat{Q}\rangle = 0$ *in a stationary state for all observables Q.*

(3) $\quad \frac{d}{dt}\langle\hat{Q}\rangle = \langle\partial\psi/\partial t|\hat{Q}|\psi\rangle + \langle\psi|\hat{Q}|\partial\psi/\partial t\rangle$

A. 1 only   B. 1 and 2 only   C. 1 and 3 only   D. 2 and 3 only   E. all of the above

Table 1 shows that only 21% of the graduate students provided the correct answer (Choice E) for question 1. Table 1 also shows that 46% and 51% of the students did not select correct statements (2) and (3) in question 1 as correct, respectively. Think-aloud interviews provided opportunities to better understand student difficulties.

Many graduate students stated that they had seen the equation in statement (1) before and recalled it from rote memory. In response to question 1, one interviewed student, who selected choice A (1 only) said, "*it is just like a definition. That's how you figure out the…how a state evolves in time is based on how it commutes with the Hamiltonian.*" Later, when explaining why he didn't choose statement (2) in question 1 as correct, he stated, "*I don't understand how the fact that the state is in a stationary state is connected to how the expectation value of an observable depends on time. What if it's not in a stationary state, what changes? I would think that it depends on the observable, because if it's in a stationary state of that observable, then….*" When the interviewer asked what he meant by "a stationary state of that observable", he replied, "*Observables have certain eigenstates, right? So if it's in an eigenstate of that observable…that's what I mean [by a stationary state of that observable].*" He then continued, "*So if it's in an eigenstate of an observable Q I would think then maybe it [the expectation value for that operator $\hat{Q}$ ] doesn't depend on time.*" Several other interviewed students also incorrectly claimed that a stationary state of an observable is an eigenstate of the operator corresponding to that observable. They did not realize that a stationary state is an eigenstate of the Hamiltonian and expectation values of all observables are time independent in a stationary state. When asked about why statement (3) is not correct, the student said, "*[statement] 3 just seems weird. It doesn't seem like something you can do…it's not a multiplication of states…you can't apply the chain rule.*"

Several other interviewees said something similar to the reasoning of the interviewed student described above, and one student even asked whether statement (3) in question 1 was designed to trick them into thinking that the chain rule for derivatives in calculus can be applied to write the partial derivative with time of the bra and ket states in an expectation value. The fact that many students explicitly noted that statement (3) did not seem technically correct but claimed that statement (1) is correct suggests that they can recall

Ehrenfest's theorem as a memorized fact but they do not understand how it is derived mathematically.

The following question was posed to investigate student difficulties with conserved quantities. A solid understanding of conserved quantities is relevant to the time dependence of expectation values.

*2. Choose all of the following statements that are <u>necessarily</u> correct.*
*(1) An observable whose corresponding time-independent operator commutes with the time-independent Hamiltonian of the system $\hat{H}$ corresponds to a conserved quantity (constant of motion).*
*(2) If an observable $Q$ does not depend explicitly on time, $Q$ is a conserved quantity.*
*(3) If a quantum system is in an eigenstate of the momentum operator, momentum is a conserved quantity.*
*A. 1 only   B. 2 only   C. 3 only   D. 1 and 3 only   E. all of the above*

Table 1 shows that 46% of students selected the correct answer (choice A) for question 2. Table 1 also shows that 18% and 42% of the graduate students incorrectly claimed that in question 2 statements (2) and (3), respectively, are correct. Individual interviews suggest that graduate students who claimed that if the system is in an eigenstate of an operator, the corresponding observable is a conserved quantity. They did not differentiate between the eigenstates of the Hamiltonian operator and those of other operators that do not commute with the Hamiltonian. During the clarification phase of the interviews, some students explicitly claimed that it does not make sense for an operator to have time dependence in its own eigenstate.

The following two questions about a spin ½ particle in an external magnetic field investigated graduate students' understanding of how expectation values of different components of the spin angular momentum evolve in time (all spin components were defined).

*For the following two questions, the Hamiltonian of a charged particle with spin-1/2 at rest in an external uniform magnetic field is $\hat{H} = -\gamma B_0 \hat{S}_z$ where the uniform field $B_0$ is along the z-direction and $\gamma$ is the gyromagnetic ratio.*
*3. Suppose the particle is initially in an eigenstate of the x component of spin angular momentum operator $\hat{S}_x$. Choose all of the following statements that are correct:*
*(1) The expectation value $\langle \hat{S}_x \rangle$ depends on time.*
*(2) The expectation value $\langle \hat{S}_y \rangle$ depends on time.*
*(3) The expectation value $\langle \hat{S}_z \rangle$ depends on time.*
*A. 1 only   B. 2 only   C. 3 only   D. 1 and 2 only   E. 2 and 3 only*
*4. Suppose the particle is initially in an eigenstate of the z component of spin angular momentum $\hat{S}_z$. Choose all of the following statements that are correct:*
*(1) The expectation value $\langle \hat{S}_x \rangle$ depends on time.*
*(2) The expectation value $\langle \hat{S}_y \rangle$ depends on time.*
*(3) The expectation value $\langle \hat{S}_z \rangle$ depends on time.*
*A. none of the above   B. 1 only   C. 2 only   D. 3 only   E. 1 and 2 only*

Table 1 shows that the correct answer choice D for question 3 was selected by 51% of the graduate students. The most common incorrect answer choice E for question 3, which was selected by 36% of the students, suggests that these students incorrectly thought that if the particle is initially in an eigenstate of the x component of spin, then the expectation value of $\hat{S}_x$ will not depend on time.

Table 1 also shows that for question 4, 33% of the students selected the correct answer choice A. Moreover, 51% of the students selected the incorrect answer choice E for question 4 suggesting they did not realize that, since the particle is initially in an eigenstate of the Hamiltonian operator, the expectation values of all observables will be time independent. In particular, in order to answer question 4 correctly, students must understand what a stationary state is and what that entails for the time dependence of expectation values of the observables. Interviews suggest that many students struggled to explain what a stationary state is.

Moreover, even those who realized that they should use the equation $\frac{d}{dt}\langle \hat{Q} \rangle = \frac{i}{\hbar} \langle [\hat{H}, \hat{Q}] \rangle$ (given in question 1) to answer both questions 3 and 4 were often not able to answer question 4 correctly because they did not understand how this equation would yield no time dependence of the expectation value of any observable in a stationary state. In order to utilize this equation to interpret that the expectation value of all observables will be time independent in a stationary state, they had to realize that the Hamiltonian in the commutator acting on the bra and ket state would give the energy corresponding to that stationary state (which is a constant) and hence the commuter becomes zero. Even during interviews, a very common incorrect answer choice for question 4 was choice E (1 and 2 only), indicating that students did not take into account the importance of the stationary state and incorrectly concluded that since $\hat{S}_x$ and $\hat{S}_y$ do not commute with the Hamiltonian, they would evolve in time.

Out of the nine interviewees who answered question 3 correctly, six answered question 4 incorrectly. When explicitly asked whether the initial conditions should matter for answering questions 3 and 4, i.e., whether it should matter whether the particle starts off in an eigenstate of $\hat{S}_x$ or $\hat{S}_z$, one student said, "*The fact that it's in an eigenstate of one or the other doesn't change anything. So it's only the expectation value of $\hat{S}_z$ that doesn't depend on time in both cases. It doesn't matter because $\hat{S}_x$ doesn't commute with the Hamiltonian. When you're finding the expectation value of $\hat{S}_x$, you sandwich it between two $|\psi(x,t)\rangle$ states. Either way you can't commute the operator that's in the middle of two exponentials that have Hamiltonians in them. [but] $\hat{S}_z$ ...you can commute it over.*" While he did reason that the commutator of the Hamiltonian and $\hat{S}_x$ was non-zero, he didn't see that the initial state was an eigenstate of the Hamiltonian and thus the system was in a stationary state and all expectation values would be time independent.

Another student stated, "*If it starts out in $\hat{S}_z$, I don't feel like that matters still, as long as the Hamiltonian still has the z dependence. I still feel like you would still have 1 and 2 true. The magnetic field would cause it to rotate in the x-y plane and have no component along z.*" This student tried to visualize what was happening but concluded that the x and y components should precess if the system started out in an eigenstate of $\hat{S}_z$. When explicitly asked whether the system was in a stationary state in question 4, he replied, "*The Hamiltonian will evolve it. Because you have $e^{-i\hat{H}t/\hbar}$ and that will evolve it in time.*". One student said that it didn't make sense that the initial states did not change his responses to questions 3 and 4, but he did not know how to make sense of the effect of the initial states mathematically. Similar to several other students, he concluded that since $\hat{S}_x$ and $\hat{S}_y$ do not commute with the Hamiltonian, their expectation values would depend on time in both questions 3 and 4 by using the equation for the time dependence of expectation value in question 1.

## SUMMARY AND FUTURE OUTLOOK

We find that many graduate students struggle with the concept of time dependence of expectation values in quantum mechanics. In particular, many students had difficulty with the two general conditions for when the expectation values of certain observables have no time dependence as discussed. Additionally, during interviews, many students struggled to describe what a stationary state of a given quantum system is. They often described a stationary state as an eigenstate of any operator corresponding to an observable. These types of difficulties suggest that while many graduate students may have developed the mathematical skills to solve complex problems in their graduate quantum mechanics classes, which often focus exclusively on quantitative facility, they still have difficulties with the basic concepts behind the mathematical manipulations. Since students are unlikely to retain what they learned without a conceptual foundation, these conceptual difficulties should be explicitly addressed in both the upper level undergraduate and graduate classes as a part of a coherent curriculum.

Based upon the research on student difficulties, we have been developing and assessing research-based learning tools to help students develop a good grasp of time dependence of the expectation value in quantum mechanics. These research-based learning tools include the Quantum Interactive Learning Tutorial (QuILT) in the context of the Larmor precession of spin, which provides a simple and intuitive platform for helping students learn about the time dependence of the expectation value of observables. The QuILT employs a guided inquiry-based approach to learning and is designed to help students build a good knowledge structure [7-9]. The instructors can use the QuILT as an in-class tutorial which students can be asked to work on in small groups of two or three and make sense of the time dependence of expectation value in the context of the Larmor precession of spin. The QuILT can also be used as a homework supplement or be utilized by underprepared graduate students as a self-study tool. We have also been developing and evaluating reflective problems which complement quantitative problems and concept tests on this topic similar to those popularized by Mazur for introductory physics courses [10]. The concept tests on this topic can be integrated with lectures and students can be encouraged to take advantage of their peers' expertise and learn from each other [10].

## ACKNOWLEDGEMENTS

We thank the National Science Foundation for awards PHY-0968891 and PHY-1202909.

## REFERENCES

1. P. Jolly, D. Zollman, S. Rebello and A. Dimitrova, Am. J. Phys. 66(1), 57 (1998).
2. C. Singh, Am. J. Phys. 69 (8), 885 (2001).
3. M. Wittmann, R. Steinberg and E. Redish, Am. J. Phys. 70(3), 218 (2002).
4. C. Singh, Am. J. Phys. 76(3), 277 (2008).
5. G. Zhu and C. Singh, Am. J. Phys. 79(5), 499 (2011).
6. C. Singh, Am. J. Phys. 76(4), 400 (2008).
7. G. Zhu and C. Singh, Phys. Rev. ST PER 8(1), 2012).
8. G. Zhu and C. Singh, Phys. Rev. ST PER 9(1), (2013).
9. G. Zhu and C. Singh, Am. J. Phys. 80(3), 252 (2012).
10. E. Mazur, Peer Instruction: A User's Manual, Prentice Hall, Upper Saddle River, NJ, 1997.